\documentclass[sigconf]{acmart}
\AtBeginDocument{%
  }

\setcopyright{acmlicensed}
\copyrightyear{2025} 
\acmYear{2025} 
\setcopyright{acmlicensed}\acmConference[KDD '25]{Proceedings of the 31st ACM SIGKDD Conference on Knowledge Discovery and Data Mining V.2}{August 3--7, 2025}{Toronto, ON, Canada}
\acmBooktitle{Proceedings of the 31st ACM SIGKDD Conference on Knowledge Discovery and Data Mining V.2 (KDD '25), August 3--7, 2025, Toronto, ON, Canada} \acmDOI{10.1145/3711896.3737230}
\acmISBN{979-8-4007-1454-2/2025/08}

\usepackage{mdframed}
\usepackage{framed}
\usepackage{enumitem}

\begin{document}

\title{GREAT: Guiding Query Generation with a Trie for Recommending Related Search about Video at Kuaishou}

% Guiding Generation with a Trie for Query Recommendation in Related Search about Videos at Kuaishou

\author{Ninglu Shao}
\affiliation{%
  \institution{Renmin University of China}
  \city{Beijing}
  \country{China}
}
\email{ninglu_shao@ruc.edu.com}

\author{Jinshan Wang}
\affiliation{%
  \institution{Kuaishou Technology Co., Ltd.}
  \city{Beijing}
  \country{China}
}
\email{wangjinshan@kuaishou.com}

\author{Chenxu Wang}
\affiliation{%
  \institution{Kuaishou Technology Co., Ltd.}
  \city{Beijing}
  \country{China}
}
\email{wangchenxu03@kuaishou.com}

\author{Qingbiao Li}
\affiliation{%
 \institution{Kuaishou Technology Co., Ltd.}
  \city{Beijing}
  \country{China}
}
\email{liqingbiao@kuaishou.com}

\author{Xiaoxue Zang}
\authornote{Corresponding author.}
\affiliation{%
  \institution{unaffiliated}
  \city{Beijing}
  \country{China}
}
\email{zxx1204007@gmail.com}

\author{Han Li}
\affiliation{%
  \institution{Kuaishou Technology Co., Ltd.}
  \city{Beijing}
  \country{China}
}
\email{lihan08@kuaishou.com}

\renewcommand{\shortauthors}{Ninglu Shao, Jinshan Wang, Chenxu Wang, Qingbiao Li \& Xiaoxue Zang}

\begin{abstract}

Currently, short video platforms have become the primary place for individuals to share experiences and obtain information. 
To better meet users' needs for acquiring information while browsing short videos, some apps have introduced a search entry at the bottom of videos, accompanied with recommended relevant queries.
This scenario is known as query recommendation in video-related search, where core task is item-to-query (I2Q) recommendation. 
As this scenario has only emerged in recent years, there is a notable scarcity of academic research and publicly available datasets in this domain.
To address this gap, we systematically examine the challenges associated with this scenario for the first time.
Subsequently, we release a large-scale dataset derived from real-world data pertaining to the query recommendation in video-\textit{\textbf{r}}elated  \textit{\textbf{s}}earch on the \textit{\textbf{Kuai}}shou app (\textbf{KuaiRS}).
Presently, existing methods rely on embeddings to calculate similarity for matching short videos with queries, lacking deep interaction between the semantic content and the query.
In this paper, we introduce a novel LLM-based framework named \textbf{GREAT}, which \textit{\textbf{g}}uides que\textit{\textbf{r}}y g\textit{\textbf{e}}ner\textit{\textbf{a}}tion with a \textit{\textbf{t}}rie to address I2Q recommendation in related search.
Specifically, we initially gather high-quality queries with high exposure and click-through rate to construct a query-based trie.
During training, we enhance the LLM's capability to generate high-quality queries using the query-based trie.
In the inference phase, the query-based trie serves as a guide for the token generation.
Finally, we further refine the relevance and literal quality between items and queries via a post-processing module.
Extensive offline and online experiments demonstrate the effectiveness of our proposed method.

\end{abstract}

\begin{CCSXML}
<ccs2012>
   <concept>
       <concept_id>10002951.10003317.10003371</concept_id>
       <concept_desc>Information systems~Specialized information retrieval</concept_desc>
       <concept_significance>300</concept_significance>
       </concept>
   <concept>
       <concept_id>10002951.10003317.10003371.10010852.10010853</concept_id>
       <concept_desc>Information systems~Web and social media search</concept_desc>
       <concept_significance>300</concept_significance>
       </concept>
 </ccs2012>
\end{CCSXML}

\ccsdesc[300]{Information systems~Specialized information retrieval}
\ccsdesc[300]{Information systems~Web and social media search}



\keywords{Large Language Model, Query Recommendation, Query Generation}


\maketitle

\begin{figure}[h]
    \centering
    \includegraphics[width=\linewidth]{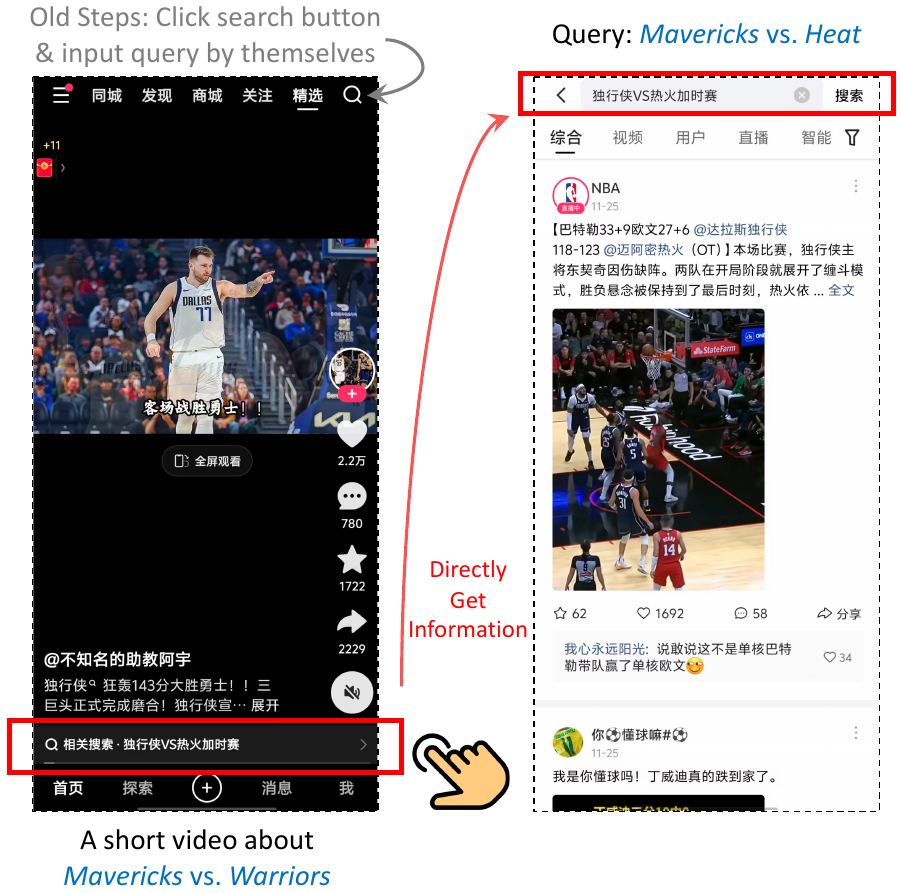}
    \caption{I2Q recommendation in related search at Kuaishou app. In recommendation scenario, users watch a short video about NBA. Then, their search intrest is triggered by the query at the bottom. After clicking the query, user enters the search result page, thus completing the transition from the recommendation scenario to the search scenario.}
    \label{fig:bottom_bar}
\end{figure}

\section{Introduction}

Nowadays, short video platforms (e.g. Kuaishou and TikTok) have become popular places for people to share their own experiences and obtain information. 
When users watch short videos, their interests are often triggered by the video content, prompting them to search for more related information~\cite{kofler2016user, si2023search}. 
As shown in Figure~\ref{fig:bottom_bar}, the user behavior chain in the above process can be described as the following steps: 
First, users are recommended an interesting short video and become interested in it. 
Next, they come up with a query and click the search button. 
Finally, they input the query and get the search results. 
Notably, there are two issues in this process: 
(1) The user behavior chain is too long, which increases the users' search cost; 
(2) The query that users come up with can be inaccurate, thus failing to meet the users' search intent.

To better meet users’ needs for acquiring information while browsing short videos, a new application scenario called query recommendation in related search has been proposed to improve the user experience.
In Figure~\ref{fig:bottom_bar}, a query related to the current video is placed at the bottom of the interface, and once users' interests are triggered by the query, they can directly click it to enter the search scenario.
This design simplifies the user behavior chain, thereby reducing the users' search cost. 
Additionally, compared to the queries that users come up with themselves, our recommended queries can more accurately describe the content (e.g., some trending topics and memes).
Moreover, it serves as a bridge between recommendation and search, seamlessly transitioning users from recommendation to search, transforming passive interests into active interests~\cite{Sun2023KuaiSARAU}. 

Although this application scenario is very important to users, academic research in this area is significantly lacking due to it being an emerging scenario and the lack of publicly available datasets. 
To bridge this gap, we thoroughly discuss the challenges of this scenario and release a large-scale and real-world dataset \textbf{KuaiRS}, which is collected from the query recommendation in \textit{\textbf{r}}elated \textit{\textbf{s}}earch scenario on the \textit{\textbf{Kuai}}shou app, a leading short-video app in China with over 400 million daily active users. 

The task of query recommendation in related search can be summarized as giving an item (e.g., short video) and then retrieving queries based on content or user behavior signals, which is known as item-to-query (I2Q) recommendation in related search. 
Unlike traditional recommendation task that primarily focus on the effectiveness (e.g., exposure and click-through rate)~\cite{covington2016deep, he2017neural}, I2Q recommendation in related search needs to consider more objectives from multiple perspectives: 
(1) Effectiveness. Similar to other recommendation tasks, user consumption metric like exposure or ctr is crucial. 
(2) Search Results Page Consumption. I2Q transfers users from recommendation to search, but if the search results page lacks relevant content, it becomes an meaningless experience for the user.
(3) Relevance. No matter how the query performs in terms of consumption, since the query is displayed to the user along with the item, an irrelevant query will disrupt the user experience.
(4) Literal Quality. Displaying queries with poor literal quality (e.g., typos and rumors) to users will also degrade user experience and can even lead to risk control or misinformation issues. 

In the industry, the most common existing methods of I2Q recommendation in related search is based on retrieval.
Retrieval-based methods take the whole item content and query as inputs separately, and use embeddings which obtained from BERT-based model to calculate similarity and recommend the query. 
However, these methods compute the embeddings of items and queries independently, which limits their effectiveness due to the lack of deep interactions between items and queries. 
Recently, Large Language Models (LLMs) have shown impressive capabilities in natural language~\cite{dubey2024llama, yang2024qwen2, fan2024survey} and multimodal tasks~\cite{yin2024survey, wang2024emu3, xue2024xgen}. 
Some methods take advantage of LLMs by employing techniques such as fine-tuning and prompt engineering to enable LLMs to directly generate queries based on items. 
Generation-based methods ensure the relevance of the query but face challenges in effectiveness and literal quality. 
Additionally, the generation-based methods are still in the early stages of exploration, offering significant promise for improvement.

Inspired by the techniques mentioned above, we introduce an innovative and effective framework called \textbf{GREAT}, which \textit{\textbf{g}}uides \textit{\textbf{q}}uery g\textit{\textbf{e}}ner\textit{\textbf{a}}tion with a \textit{\textbf{t}}rie to address the challenges faced by generation-based methods. 
Leveraging LLMs, \textbf{GREAT} retains the benefits of generation-based methods in terms of relevance, while tackling issues of effectiveness and literal quality by incorporating high-quality query data during both training and inference. 
Specifically, we begin by collecting high-quality queries that have high exposure and click-through rate, and then construct a query-based trie from these queries.
Since queries generated directly by LLMs often suffer from literal quality problem (e.g., typos and rumors), we use the query-based trie to guide the generation results of LLMs. 
During training, we incorporate the Next-Token in Trie Prediction (NTTP) task using the query-based trie to enhance the LLM's ability to generate high-quality queries. 
During inference, the query-based trie guides the generation of the next token, thereby improving the quality of the generated queries.
Furthermore, to further ensure the relevance and literal quality of the queries, a post-processing module is added during inference to filter out low-quality content.

To summarize, the following contributions are highlighted in for our paper:
\begin{itemize}[leftmargin=*]
    \item To the best of our knowledge, this is the first systematic discussion of I2Q recommendation in related search. 
    We clarify the four main challenges, which consists of effectiveness, search results page consumption, relevance and literal quality.
    \item There are no publicly available datasets in this field, which greatly hinders the development of academic research.
    To bridge this gap, we release an large-scale and real-world dataset \textbf{KuaiRS}, which is collected from Kuaishou, a leading short-video app in China with over 400 million daily active users.
    \item We propose a novel framework \textbf{GREAT} which utilizes LLMs to address the three challenges in I2Q recommendation. 
    During both training and inference, \textbf{GREAT} takes advantage of query-based trie to enhance the capability of LLMs. 
    \item Extensive offline experiments demonstrate the effectiveness of \textbf{GREAT}. 
    Moreover, the performance of \textbf{GREAT} in online industrial scenario of Kuaishou further highlights its powerful capability.
\end{itemize}

\section{Related Work}

\subsection{Retrieval-based Method in I2Q}

Although I2Q recommendation in related search is a relatively new scenario, it has greatly improved the fulfillment of users' information acquisition needs.
Its core task is to recommend one or more queries based on the target item.
Therefore, many retrieval-based methods can be adapted to this new scenario.
These methods either pre-build the I2Q index~\cite{Yang2020LargeSP} or employ approximate k-nearest neighbor approach~\cite{Johnson2017BillionScaleSS} to retrieve relevant queries online.
The former typically depends on a substantial amount of collaborative signals from user behavior~\cite{Yang2020LargeSP, Zhu2018LearningTD}.
However, due to the lack of item-query interactions, these methods cannot address the cold start problem for items and queries~\cite{Wei2021ContrastiveLF, Lee2019MeLUMU}.
To address this issue, some content-based retrieval methods have been introduced.
They perform retrieval by evaluating the content similarity between items and queries.
One type of method relies on term-level lexical matching to calculate the similarity between items and queries, thereby providing better interpretability~\cite{Ramos2003UsingTT, Robertson2009ThePR}.
Another type of method uses neural networks to transform items and queries into embeddings in the same latent space~\cite{Chen2024BGEMM, radford2021learning, li2022blip}, subsequently designing a two-tower structure to calculate similarity~\cite{Xiao2022ProgressivelyOB, Xiao2021TrainingLN}.
However, due to the inherent limitation of the two-tower structure, these methods lack deep interactions between items and queries within the network, lowering the upper bound of effectiveness~\cite{Chang2020PretrainingTF, Haldar2020ImprovingDL}.

\subsection{Generation-based Method in I2Q}

In recent years, LLMs have become a significant research focus in both academia and industry~\cite{achiam2023gpt, dubey2024llama, zhu2024yulan}.
Owing to their immense potential capabilities, LLMs have been introduced into various application scenarios, such as recommendation~\cite{dai2023uncovering, Zhai2024ActionsSL, Jia2024KnowledgeAF} and time series forecasting~\cite{Jin2024TimeSF}. 
In the context of I2Q recommendation in related search, the generation-based approach directly generates queries related to the item, thereby avoiding the need to retrieve query from query pool and improving the relevance between the query and the item.
There are two types of generation-based methods in I2Q recommendation in related search.
The first type of methods directly utilizes off-the-shelf LLMs, typically employing prompt engineering or framework design to generate queries based on item content~\cite{White2023APP, Sahoo2024ASS}. 
However, these methods heavily rely on the design of prompts or frameworks, and the generated query style often does not align with the platform's style, which results in poor performance in online industrial scenario.
The second type of methods fine-tunes LLMs to customize the style and format of the query~\cite{Parthasarathy2024TheUG, zhang2023instruction}. 
However, in practice, we find that due to the lack of high-quality domain data, the generated queries often have poor literal quality, significantly degrading user experience.

\section{Dataset Description}

\begin{table}[h]
    \centering
    \begin{tabular}{lr|lr}
    \toprule
        \multicolumn{4}{c}{Training Dataset} \\
    \midrule
        \#video-query pair & 1M & avg. \#words per caption & 32.55 \\
        \#video & 439,896 & avg. \#words per ocr\_cover & 16.24 \\
        \#query & 419,238 & avg. \#words per query & 7.41 \\
    \midrule[0.75pt]
        \multicolumn{4}{c}{Validation Dataset} \\
    \midrule
        \#video-query pair & 10K & avg. \#words per caption & 35.15 \\
        \#video & 7,814 & avg. \#words per ocr\_cover & 17.74 \\
        \#query & 7,965 & avg. \#words per query & 6.98 \\
    \midrule[0.75pt]
        \multicolumn{4}{c}{Test Dataset} \\
    \midrule
        \#video-query pair & 10K & avg. \#words per caption & 34.85 \\
        \#video & 7,791 & avg. \#words per ocr\_cover & 17.70 \\
        \#query & 7,955 & avg. \#words per query & 7.03 \\
    \bottomrule
    \end{tabular}
    \caption{Basic statistics of KuaiRS.}
    \label{tab:dataset_statistics}
\end{table}

In this section, we present the large-scale, real-world dataset named KuaiRS.
We begin by describing the general overview of KuaiRS, including its data sources and fundamental structure.
Following this, we introduce the data construction workflow of KuaiRS, including data collection and data cleaning.
Finally, we provide some statistical information about KuaiRS.

\subsection{Dataset Overview}

To address the lack of publicly available datasets for I2Q recommendations in related search, we release KuaiRS, which is derived from the Kuaishou app.
Kuaishou is one of the most popular short video platforms in China, with over 400 million daily active users.
As shown in Figure~\ref{fig:bottom_bar}, Kuaishou provides I2Q recommendation in related search service. 
When users watch the short video, they can click the query at the bottom of interface instead of manually inputting query to search. 
After clicking the query, they directly enter the serach results page to find short videos of interest.

The data of KuaiRS is divided into two parts: video part and query part.
The video part comprises the video caption (labeled as 'caption') and the content of the video cover (labeled as `ocr\_cover').
The former refers to the caption created by the user during the upload of the short video. 
The latter  pertains to the text content of the video cover uploaded by the user, identified through OCR technology.
The query part contains only one field, which is the query itself (labeled as 'query').

\subsection{Data Construction}

To facilitate the research on I2Q recommendation in related search, KuaiRS is constructed with the following steps:

First, we collect user logs from the Kuaishou app between May 23, 2024 and May 29, 2024.
These user logs are aggregated by item and query to obtain video-query pairs.
Each video-query pair contains two types of information.
One type is the metadata of the video and query, including video caption, content of the video cover, query.
The other type is the side information used for subsequent data cleaning, including exposure, CTR, and so on.

Second, we perform data cleaning using various side information.
In order to improve quality of the dataset, we filter out data with less than 1,000 exposures or less than 10 clicks.
To further ensure the relevance between item and query, we use MBVR~\cite{wang2022modality} to calculate the similarity between short video and query, and set a threshold of 0.44 based on experience for filtering.
Additionally, considering data ethics and safety issues, data containing sensitive words is also filtered.

Finally, we randomly sample 1.02 million video-query pairs from the cleaned data.
To ensure consistency with the online industrial scenario, we sort the data chronologically and divide it into two parts.
The first part is the training set, which contains the first 1 million entries of the data.
The remaining part is divided into the validation set and test set, with 20,000 entries randomly split into 10,000 each, ensuring consistent distribution between the validation and test sets.

\subsection{Statistics}

KuaiRS contains real records of 1.02 million video-query pairs collected over one week on the Kuaishou app.
The fundamental statistics of KuaiRS are summarized in Table 1.
Overall, the average length of captions in the training set, validation set, and test set exceeds 30, the average length of `ocr\_cover' is approximately 17, and the average length of `query' is about 7.
Moreover, the training set comprises 439,896 short videos, which means that on average, each video corresponds to approximately 2 queries as ground truth.
Similarly, each video in the validation set and test set corresponds to about 1 query as ground truth.
For more specific statistics and usage, please refer to \href{https://github.com/rainym00d/KuaiRS}{https://github.com/rainym00d/KuaiRS}.

\begin{figure*}[t]
    \centering
    \includegraphics[width=\linewidth]{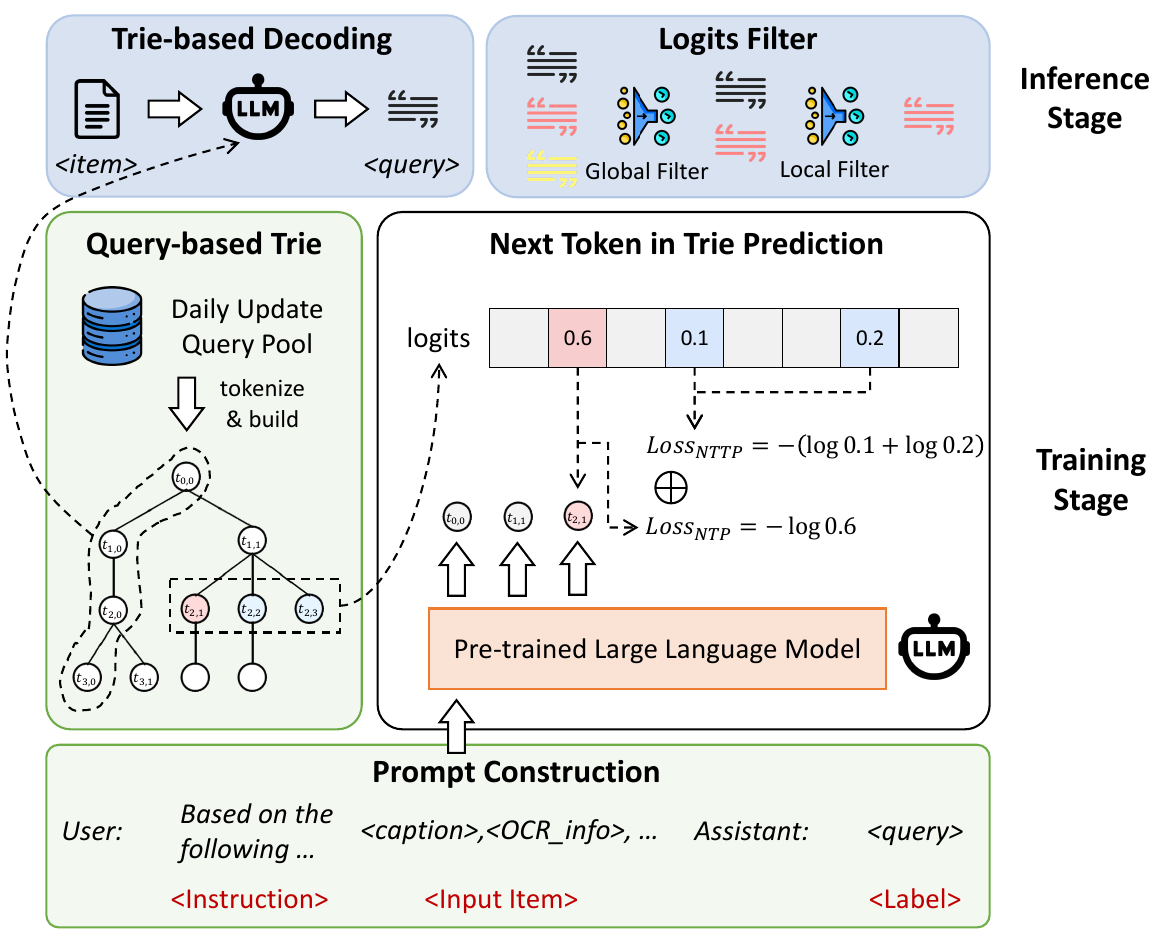}
    \caption{The framework of GREAT. During the data preparation, we construct prompts for LLMs' inputs and build a query-based trie by collecting queries with high exposure and click-through rate. During training, the NTTP task is introduced to improve the capability of LLMs in generating high-quality queries. During inference, Trie-based Decoding guides LLMs to generate the next token, reducing hallucinations and rumors in queries,  while the Logits Filter further filters out low-quality results, ensuring high relevance and literal quality of queries.}
    % \caption{The framework of GREAT.}
    \label{fig:framework}
\end{figure*}

\section{Method}

In this section, we delve into the technical aspects of GREAT.
We first describe the architecture of GREAT.
Subsequently, we detail the data preparation for GREAT, including prompt construction and query-based trie.
Then, we introduce the Next Token Prediction in Trie task during the training, which facilitates the LLMs generate high-quality queries.
Finally, we discuss the techniques in the inference, focusing on two basic components: Trie-based Decoding and Logits Filter.

\subsection{Framework of GREAT}

As illustrated in Figure~\ref{fig:framework}, GREAT consists of five modules: prompt construction, query-based trie, NTTP, Trie-based Decoding, and Logits Filter.
During data preparation, we construct prompts based on short video data to serve as input for LLMs.
Meanwhile, high-quality queries with high exposure and click-through rate are collected to build a query-based trie, which is updated on a daily basis. 
During training, we employ the query-based trie to add the NTTP task, thereby assisting LLMs in generating high-quality queries.
During inference, Trie-based Decoding is used to guide LLMs to generate the next token along the path of the query-based trie, thus reducing hallucinations and rumors. 
Furthermore, the Logits Filter module performs post-processing on queries from both global and local viewpoints, filtering out low-quality queries to further guarantee the literal quality of the queries.

\subsection{Data Preparation}

\subsubsection{Prompt Construction}

We design a prompt and incorporat the content of short videos into it.
To leverage the generative capabilities of LLMs, we use the prompt as input to LLMs to generate queries, thereby achieving the I2Q recommendation in related search tasks.
Specifically, we propose the following prompt template for I2Q recommendation in related search:

\begin{mdframed}
    \textbf{User:} <Instruction>\textbackslash n<Input Item>\textbackslash n\textbf{Assistant:} <Output>
\end{mdframed}

In this template, both `User' and `Assistant' serve as the original chat templates for LLMs, and different LLMs may have different references~\cite{yang2024qwen2}. 
For instance, certain LLMs use `[INST]' to denote `User' and `[/INST]' to denote `Assistant' in their chat templates~\cite{touvron2023llama}. 
Moreover, <Instruction>, <Input Item> and <Output> act as placeholders, requiring the insertion of appropriate content. 
During practical application, the template is populated with the short video's content, including `caption' and `ocr\_info':

\begin{mdframed}
    \textbf{User:} Based on the content of the following video, generate one keyword that the user might be interested in. \\
    \textit{<caption>}, \textit{<ocr\_info>}, \dots \\
    \textbf{Assistant:} \textit{<query>}
\end{mdframed}

As shown at the bottom of the Figure~\ref{fig:framework}, the completed prompt is tokenized and fed into the LLM.
Subsequently, the LLM generate suitable query according to the content of the short video.

\subsubsection{Query-based Trie}

The query-based trie is one of the core components of GREAT, playing a significant role in both the training and inference phases of the model. 
As illustrated on the left side of Figure~\ref{fig:framework}, the query-based trie not only deeply participates in the model's training but also guides the decoding strategy during inference, helping the LLM generate high-quality queries. 
The primary aim of this design is to enhance the effectiveness of queries generated by LLMs through the guidance provided by the query-based trie. 
Consequently, the quality of queries within the query-based trie directly affects the overall performance of GREAT.

In particular, we collect real-world data from user logs and integrate it with user behavior data to select queries with high exposure and high click-through rate, ensuring that the query pool is both comprehensive and representative of actual user needs.
Subsequently, we conduct a thorough review of the preliminary filtered queries, removing those that has literal quality issues (e.g., typos or rumors), to ensure the literal quality of the query pool.
Additionally, to address the dynamic changes in user needs, we implement a daily update mechanism to continuously introduce the latest query data, ensuring the timeliness of the query pool.
Considering the limitations of storage and computational resources in practical applications, we define a time window (e.g., 15 days) for the query pool to control its scale.
Finally, we employ the tokenizer of LLMs to convert the queries in the query pool from text into tokens, and then construct a trie using these tokens as nodes.

\subsection{NTTP Task in Training}

Next Token Prediction (NTP) serves as the fundamental task in training and inference for autoregressive LLMs~\cite{Vaswani2017AttentionIA, achiam2023gpt}.
Whether through unsupervised pretraining or instruction tuning, LLMs gain new knowledge by adjusting the distribution in the latent space to accurately predict the next token.
Nonetheless, in the I2Q recommendation task in related search, depending solely on NTP to learn the semantic relationship between items and queries is insufficient.
In practical applications, we find that directly using off-the-shelf LLMs or fine-tuned LLMs to generate queries results in significant risk control and literal quality issues, leading to high costs in online industrial scenarios.

To address this issue, we introduce the Next Token in Trie Prediction (NTTP) task as an auxiliary component during the training of GREAT.
The NTTP task allows LLMs to learn high-quality queries from the query-based trie, enhancing the alignment of the sample space for training and inference and boosting the model's performance in practical applications.
With the introduction of the NTTP task in training, LLMs are required to predict the next token in the labels while also paying attention to the next token in the trie.
This mechanism drives LLMs to acquire the knowledge of high-quality queries in the query-based trie.
Specifically, we represent the current sequence as $s_{i-1}: t_0, t_1, \dots, t_{i-1}$ and predict the next token $t_i$.
Next, we locate the set of child nodes at the next level of the path $s_{i-1}$ in the query-based trie, denoted as $\text{Trie}(s_{i-1})$.
The loss function for the NTTP task is defined as follows:
\begin{equation*}
    \mathcal{L}_{NTTP} = \sum\nolimits_{t_j \in \text{Trie}(s_{i-1})} log(p_{t_j}),
\end{equation*}
where $\mathcal{L}_{NTTP}$ denotes the loss of the NTTP task, $t_j$ is a token in the set $\text{Trie}(s_{i-1})$, and $p_{t_j}$ is the probability of the model output $t_j$.
Therefore, the complete loss function for $t_i$ is defined as follows:
\begin{equation*}
    \mathcal{L} = \mathcal{L}_{NTP} + \alpha \cdot \mathcal{L}_{NTTP},
\end{equation*}
where $\mathcal{L}_{NTP}$ represents the loss of NTP task, and $\alpha$ is the hyperparameter that controlling the weight of NTTP task.

\subsection{Techniques in Inference}

\subsubsection{Trie-based Decoding}

During the inference phase, beam search or greedy search is typically employed to directly generate queries.
However, in online industrial scenario, we find that the queries generated by these methods, while performing well in terms of relevance, exhibit significant flaws in literal quality.
Specifically, these queries overly focus on relevance and neglect the authenticity of the content.
In practical applications, although these low-quality cases can be filtered out through other modules, this approach does not fundamentally address the issue and instead leads to the unnecessary consumption of computational resources.

To address this issue, we propose Trie-based Decoding to assist the inference of LLMs.
By introducing a query-based trie, this method guides the generation of the next token, thus integrating high-quality query information into the generation and significantly improving the literal quality of the query.
In particular, Trie-based Decoding avoids utilizing the full vocabulary table and instead focuses on the child nodes of the current sequence in the query-based trie.
As illustrated in the left part of Figure~\ref{fig:framework}, suppose the current sequence is $t_{0,0}, t_{1, 0}$.
Since $t_{1, 0}$ has only one child node $t_{2, 0}$, the LLM's candidate set is limited to $t_{2, 0}$, thereby updating the current sequence to $t_{0,0}, t_{1, 0}, t_{2, 0}$.
Similarly, when the sequence reaches $t_{2, 0}$, since it has two child nodes, $t_{3, 0}$ and $t_{3, 1}$, the LLM will select the next token based on their logits values.
In this example, the LLM selects $t_{3, 0}$ as the next token. 
Finally, the generated query sequence is $t_{0,0}, t_{1, 0}, t_{2, 0}, t_{3, 0}$.

\subsubsection{Logits Filter}

We develop a post-processing module named Logits Filter to assess the quality of generated queries.
The core idea of the Logits Filter is to utilize the softmax-normalized logits values output by LLMs as a basis for measuring the relevance and literal quality of queries.
For each candidate token $t_i$, its score $p_{t_i} = P(t_i|I, s_{i-1})$ indicates the likelihood of the token being chosen given the current item $I$ and context $s_{i-1}$.
By leveraging this probability distribution, the Logits Filter can select the token sequence that best matches the item, thus generating high-quality queries.
This mechanism is based on a key assumption that well-trained LLMs have incorporated the knowledge pertaining to I2Q recommendation in related search, enabling them to accurately select the next token.

The Logits Filter is composed of two components: the Global Filter and the Local Filter.
The Global Filter assesses the quality of a query as a whole. 
For a query $q: t_0, t_1, \dots, t_n$, its global score is defined as:
\begin{equation*}  
    F_{G}(q) = \frac{1}{n} \sum_{i=0}^{n} p_{t_i},  
\end{equation*}  
where $n$ is the length of the query after tokenization.
By applying the Global Filter, queries from the query set $Q$ that meet the condition $F_{G}(q) > \theta_{G}\}$ are selected, yielding $Q_{G}=\{q \in Q | F_{G}(q) > \theta_{G}\}$, where $\theta_{G}$ being the threshold hyperparameter of the Global Filter.
The Local Filter focuses on the quality of individual tokens, and its calculation method is as follows:
\begin{equation*}  
    F_{L}(q) = \min(p_{t_0}, p_{t_1}, \dots, p_{t_n}).  
\end{equation*}  
Similarly, we further filter out queries from $Q_{G}$ that satisfy $F_{L}(q) > \theta_{L}\}$, resulting in $Q_{L}=\{q \in Q_{G} | F_{L}(q) > \theta_{L}\}$, where $\theta_{L}$ is the threshold hyperparameter of the Local Filter.

\section{Experiments}

\subsection{Experiment Setting}

\subsubsection{Implementations}

In this study, we select Qwen 2.5 1.5B~\cite{yang2024qwen2} as the base model, which achieves a good balance between performance and scale, making it especially appropriate for online industrial scenario applications.
In the offline experiments, we train and evaluate GREAT and other baselines on the KuaiRS dataset, with all queries from KuaiRS serving as the query pool.
The experimental environment is configured with a machine equipped with Nvidia 8 $\times$ V100 GPU.
In training, the learning rate is set to 5e-5, the batch size is 128\footnote{The batch size for each GPU is 4, and we set gradient accumulation to 4. Therefore, in Distributed Data Parallel training with 8 GPUs, the total batch size is 128.}, and the weight hyperparameter $\alpha$ for NTTP loss is set to 0.1.
To ensure the comparability of recall quantities across different methods in the offline experiments, the Logits Filter is not employed.
In the online experiments, the Logits Filter's threshold hyperparameters $\theta_{G}$ and $\theta_{L}$ are set to 0.2 and 0.05, respectively.

\subsubsection{Metrics}

To assess the offline performance of I2Q recommendation in related search, we use the similarity between the retrieved results and the ground truth as the evaluation criterion.
Traditional retrieval models can directly compare their results with the ground truth since the results are necessarily within the query pool.
However, the results of generative models may extend beyond the query pool, making it difficult to directly compare them with the ground truth.
To address this issue, we propose a new evaluation metric called Edit@k.
Edit@k evaluates performance by computing the average edit distance between the top-k retrieved results and the ground truth, where a smaller value indicates better performance.
The definition of Edit@k is as follows:
\begin{equation*}
    \text{Edit@k} = \frac{1}{k} \sum_{i=1}^{k} \text{edit\_distance}(q, \hat{q}_i),
\end{equation*}
where $q$ represents the ground truth, and $\hat{q}_i$ denotes the $i$-th retrieved query.

In the online experiments, we evaluate the model's performance from the following three dimensions: effectiveness, relevance, and literal quality.
Firstly, the model's effectiveness is primarily measured by user consumption behaviors, including user consumption of queries and user consumption on search result pages. 
We use metrics such as exposure, click-through rate (CTR), and CTR on search results page  to quantify the model's effectiveness.
Secondly, we analyze the relevance between item and query through human evaluation.
In detail, results randomly sampled from the online industrial scenario are manually labeled as either relevant or irrelevant, and the percentage of relevant results is then calculated.
Finally, the evaluation of the literal quality focuses on detecting quality issues such as grammatical errors, typos, or rumors within the queries.
Similar to the relevance assessment, we also employ human evaluation to label the queries and calculate the proportion of queries without literal quality issues.

\subsection{Offline Experiments}

\begin{table*}[h]
    \centering
    \begin{tabular}{l|ccccc}
    \toprule
        Method & Edit@1 & Edit@5 & Edit@10 & Edit@20 & avg. \\
    \midrule
        SimCSE & 4.81 & 5.37 & 5.69 & 6.01 & 5.47 \\
        bge-base-zh-v1.5 & 5.43 & 5.97 & 6.27 & 6.59 & 6.07 \\
        bge-large-zh-v1.5 & 5.37 & 5.92 & 6.23 & 6.54 & 6.02 \\
        bge-base-zh-v1.5 fine-tuned & 4.79 & 5.41 & 5.67 & 5.97 & 5.46 \\
        bge-large-zh-v1.5 fine-tuned & 4.71 & 5.35 & 5.62 & 5.94 & 5.41 \\
    \midrule
        Qwen 2.5 1.5B zero shot & 9.37 & 15.24 & 17.89 & 19.92 & 15.60 \\
        Qwen 2.5 1.5B fine-tuned & 4.48 & 5.28 & 5.73 & 6.10 & 5.40 \\
    \midrule
        GREAT (Ours) & \textbf{4.34} & \textbf{5.20} & \textbf{5.53} & \textbf{5.80} & \textbf{5.22} \\
    \bottomrule
    \end{tabular}
    \caption{The performance of different methods in offline experiments.}
    \label{tab:offline}
\end{table*}

We first make offline experiments to demostrate the effectivenees of GREAT. 
We make comparasion with various baselines, including traditional retrieval-based and emerging generative-based methods. 

\begin{itemize}[leftmargin=*]
    \item \textbf{SimCSE} utilizes contrastive learning to train embeddings, demonstrating excellent performance in many retrieval tasks.
    \item \textbf{BGE} is the popular model on the MTEB leaderboard.
    It shows powerful performance and offers various options of model.
    \item \textbf{Qwen 2.5} is a commonly used generative model. In our experiments, we evaluate its zero-shot and fine-tuned performance.
\end{itemize}

The results are shown in Table~\ref{tab:offline}. The following observations can be drawn from it. 
Firstly, GREAT performs the best among all methods. 
By comparison, it can be found that the fine-tuned generation-based methods significantly outperform the traditional retrieval-based methods in the I2Q recommendation in related search task. 
Specifically, both GREAT and the fine-tuned Qwen 2.5 1.5B are better than SimCSE and BGE. 
Compared to retrieval-based methods, large models possess strong content understanding capability, thereby expanding the boundaries of the I2Q recommendation in related search task. 
Furthermore, GREAT surpasses the fine-tuned Qwen 2.5 1.5B, indicating that NTTP and Trie-based decoding techniques can further improve the performance of LLMs in the I2Q recommendation in related search task. 
Secondly, by comparing BGE models with different sizes, we find that models with larger parameters perform better. 
Specifically, compared to the base version of BGE, the larger version of BGE delivers better results, which is also validated in the fine-tuned BGE models. 
This phenomenon suggests that, when computational resources are sufficient, using larger models may lead to more significant improvements. 
Thirdly, whether retrieval-based or generation-based methods, fine-tuning can enhance performance, especially in generation-based methods. 
By examining the generated cases, we found that most of the queries produced by Qwen 2.5 1.5B in zero-shot mode do not meet our formal requirements, resulting in particularly poor performance.
This indicates that if one intends to use off-the-shelf LLMs for the I2Q recommendation in related search task, instruction following will be a challenge.

\subsection{Online Experiments}

\begin{table*}[t]
    \centering
    \begin{tabular}{l|ccc|c|c}
    \toprule
        Method & Exposure & CTR & $\text{CTR}_{SRP}$ & Relevance & Literal Quality \\
    \midrule
        SimCSE & +0.062\% & +0.027\% & -0.053\% & +1.0\% & -0.5\% \\
        bge-large-zh-v1.5 fine-tuned & +0.127\% & +0.079\% & -0.088\% & +2.5\% & -1.5\% \\
    \midrule
        Qwen 2.5 1.5B fine-tuned & -0.158\% & +0.109\% & -0.146\% &\textbf{ +7.5\%} & -5.0\% \\
    \midrule
        GREAT (Ours) & \textbf{+0.251\%} & \textbf{+0.174\%} & \textbf{+0.396\%} & +5.5\% & \textbf{+6.0\%} \\
    \bottomrule
    \end{tabular}
    \caption{The performance of different methods in online A/B test. ($\text{CTR}_{SRP}$ stands for CTR on search results page.)}
    %\caption{The performance of different methods in online A/B test on Kuaishou app. ($\text{CTR}_{SRP}$ stands for CTR on search results page.)}
    \label{tab:online}
\end{table*}

\begin{figure}[h]
    \centering
    \includegraphics[width=\linewidth]{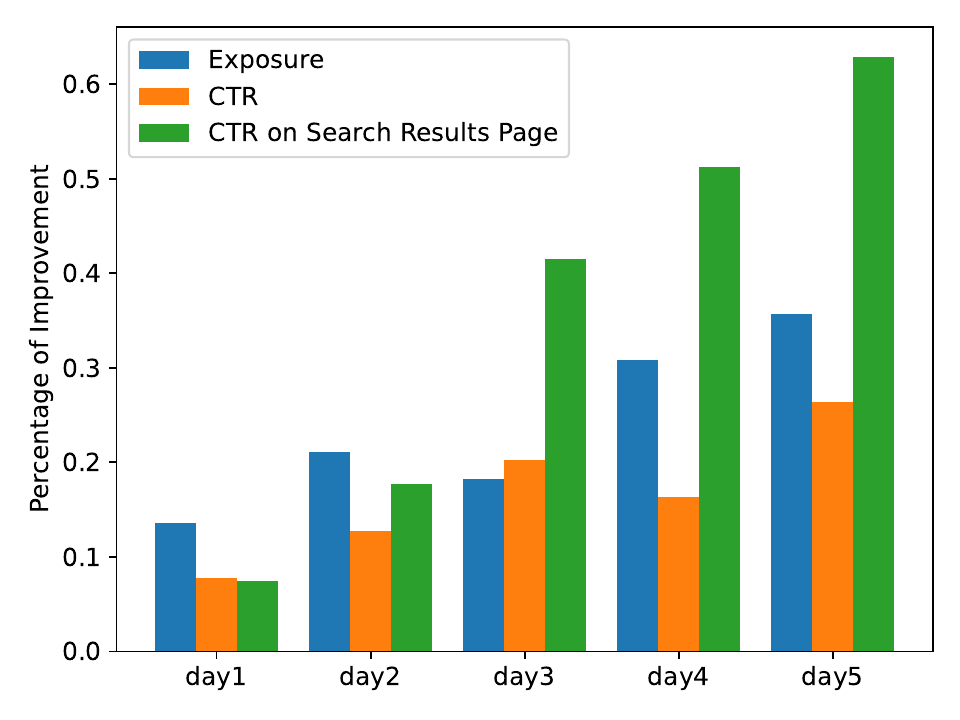}
    \caption{The performance of GREAT in online A/B test on Kuaishou app over 5 consecutive days}
    \label{fig:online}
\end{figure}

To evaluate the effectiveness of GREAT in online industrial scenarios, we deploy GREAT on the Kuaishou app and conduct an online A/B test over a period of 5 days. 
Due to data security concerns, all results are presented in relative values.

The results are shown in Tbale~\ref{tab:online}, where the following observations can be made.
Firstly, GREAT's average exposure, CTR, and CTR on the search results page are all higher than other baselines, with values of +0.251\%, +0.174\%, and +0.396\%, respectively.
This indicates that our method can better model user consumption preferences.
Secondly, we conduct evaluations of relevance and literal quality.
We randomly select 200 exposure data from both the control group and the experimental group for human evaluation.
Compared to the online baseline, our method improves relevance and literal quality by 5.5\% and 6.0\%, respectively.
It is worth noting that the fine-tuned Qwen 2.5 1.5B model, while ranking first in relevance, focuses excessively on relevance, which compromises its performance on other metrics.
Finally, compared to other baselines, GREAT demonstrates improvements across all metrics, avoiding the seesaw phenomenon\cite{Tang2020ProgressiveLE}.
This finding suggests that GREAT can effectively achieve multi-objective optimization, enhancing overall performance while maintaining metric balance.

As illustrated in Figure~\ref{fig:online}, we performe a day-level detailed statistical analysis of GREAT in online A/B test on Kuaishou app, leading to the following two primary conclusions.
Firstly, GREAT consistently outperform the online baseline in terms of exposure, CTR, and CTR on the search results page daily, without significant fluctuations.
This phenomenon indicates that our method not only achieves significant performance improvements but also exhibits excellent stability.
Secondly, by observing the 5-day experimental period, it can be seen that all evaluation metrics show a steady upward trend overall, reaching their peak on the last day of the experiment.
This trend is primarily attributed to the increasing accumulation of data over time, which allows the queries generated by GREAT to gain more online exposure, thereby achieving continuous performance enhancement.

\begin{table}[h]
    \centering
    \begin{tabular}{l|cccc}
    \toprule
        Setting & Edit@1 & Edit@5 & Edit@10 & Edit@20 \\
    \midrule
        GREAT & \textbf{4.34} & \textbf{5.20} & \textbf{5.53} & \textbf{5.80} \\
        w/o NTTP & 4.42 & 5.25 & 5.68 & 5.97 \\
        w/o TD & 4.37 & 5.24 & 5.62 & 5.93 \\
        w/o Both & 4.48 & 5.28 & 5.73 & 6.10 \\
    \bottomrule
    \end{tabular}
    \caption{Ablation study of GREAT in effectiveness. The experiments are performed on test dataset of KuaiRS. (w/o TD stands for GREAT without Trie-based Decoding.)}
    \label{tab:ablation_offline}
\end{table}

\begin{table}[h]
    \centering
    \begin{tabular}{l|cc}
    \toprule
        Setting & Relevance & Literal Quality \\
    \midrule
        GREAT & +6.22\% & \textbf{+6.91\%} \\
        w/o NTTP & +5.04\% & +3.03\% \\
        w/o Trie-based Decoding & \textbf{+7.21\%} & +0.4\% \\
        w/o Logits Filter & +1.68\% & +2.8\% \\
    \bottomrule
    \end{tabular}
    \caption{Ablation study in relevance and literal quality.}
    %\caption{Ablation study of GREAT in relevance and literal quality.}
    \label{tab:ablation_online}
\end{table}

\subsection{Ablation Study}

In this subsection, we conduct ablation study to analyze the influential factors of GREAT, whose results are presented with Table~\ref{tab:ablation_offline} and~\ref{tab:ablation_online}.

Firstly, we explore the impact of NTTP and Trie-based Decoding on GREAT in terms of effectiveness. 
\textbf{w/o NTTP} denotes that the query-based trie is only used during the inference stage, without incorporating user consumption preference information for queries during training. 
\textbf{w/o Trie-based Decoding} means that user consumption preference information for queries is included during training, but only beam search is used during decoding. 
\textbf{w/o Both} refers to basic fine-tune of LLMs. 
On the one hand, both w/o NTTP and w/o Trie-based Decoding outperform w/o Both, indicating that these two modules are effective when used independently. 
On the other hand, GREAT surpasses all its variants, suggesting that its performance can be further enhanced when these modules are used together.

Secondly, we investigate the effects of NTTP, Trie-based Decoding, and Logits Filter on GREAT in terms of Relevant and Literal Quality. 
To ensure fairness in experiments, we randomly sample 100 short videos as items, keeping other settings consistent with online industrial scenario evaluations. 
For data security reasons, the data in Table~\ref{tab:ablation_online} are presented as relative values, using the LLMs with only fine-tuning as the baseline. 
\textbf{w/o Logits Filter} denotes the post-processing module is not used. 
Overall, GREAT and all its variants outperform the baseline, which indicates that all our modules have a positive impact on both Relevant and Literal Quality. 
Moreover, GREAT surpasses both w/o NTTP and w/o Logits Filter in terms of Relevant and Literal Quality, proving the necessity of incorporating user consumption preference information during training and adding post-processing in inference stage. 
It is worth noting that w/o Trie-based Decoding performs better than GREAT in terms of Relevant Quality but significantly worse in Literal Quality compared to other variants. 
We believe that without the constraints of query-based trie, LLMs generate more queries that are relevant but lower literal quality.

\section{Conclusion}

In this paper, we systematically analyze the I2Q recommendation in related search for the first time.
To tackle the scarcity of publicly available datasets in this field, we release a large-scale, real-world dataset named KuaiRS, collected from the Kuaishou app.
Moreover, we propose a novel LLM-based approach called GREAT for I2Q recommendation in related search.
GREAT consists of five key components: prompt construction, query-based trie, NTTP, Trie-based Decoding, and Logits Filter. 
During the data preparation, we construct prompts for LLMs' inputs and build a query-based trie by collecting queries with high exposure and click-through rate.
During training, the NTTP task is introduced to improve the capability of LLMs in generating high-quality queries.
During inference, Trie-based Decoding guides LLMs to generate the next token, reducing hallucinations and rumors in queries,  while the Logits Filter further filters out low-quality results, ensuring high relevance and literal quality of queries.
Extensive offline and online experiments demonstrate the effectiveness of GREAT.
Additionally, the results show that GREAT not only increases user consumption of queries but also maintains the relevance and literal quality.

\section{Acknowledgments}

This work was funded by Kuaishou Technology.
We also thank Han Li, for his strong support and invaluable guidance in this project.

\clearpage

% \begin{acks}
% \end{acks}

\bibliographystyle{ACM-Reference-Format}
\balance
\bibliography{ref}

\appendix

\end{document}